\documentclass[british,english,12pt]{iopart}
\usepackage[T1]{fontenc}
\usepackage[latin9]{inputenc}
\usepackage{babel}
\usepackage{amsbsy}
\usepackage{amstext}
\usepackage{amssymb}
\usepackage{graphicx}
\usepackage{esint}
\usepackage{subscript}
\usepackage[unicode=true,
 bookmarks=true,bookmarksnumbered=false,bookmarksopen=false,
 breaklinks=true,pdfborder={0 0 0},backref=false,colorlinks=false]
 {hyperref}
\hypersetup{pdftitle={Large momentum-dependence of the main dispersion kink in the high-Tc superconductor Bi2Sr2CaCu2O8+d},
 pdfauthor={N. C. Plumb et al.}}

\makeatletter
\usepackage{iopams}
\usepackage{setstack}

\makeatother

\begin{document}
\title[Large momentum-dependence of the main ``kink'' in Bi$_2$Sr$_2$CaCu$_2$O$_{8+\delta}$]{Large momentum-dependence of the main dispersion ``kink'' in the high-$T_c$ superconductor Bi$_2$Sr$_2$CaCu$_2$O$_{8+\delta}$}

\author{N. C. Plumb$^{1}$\footnote{Present address: Swiss Light Source, Paul Scherrer Institut, CH-5232 Villigen PSI, Switzerland},
T. J. Reber$^{1}$, H. Iwasawa$^{2}$, Y. Cao$^{1}$, M. Arita$^{2}$,
K. Shimada$^{2}$, H. Namatame$^{2}$, M. Taniguchi$^{2}$, Y. Yoshida$^{3}$,
H. Eisaki$^{3}$, Y. Aiura$^{3}$ and D. S. Dessau$^{1,4}$}

\address{$^{1}$ Department of Physics, University of Colorado, Boulder, CO
80309-0390, USA}

\address{$^{2}$ Hiroshima Synchrotron Radiation Center, Hiroshima University,
Higashi-Hiroshima 739-0046, Japan}

\address{$^{3}$ National Institute of Advanced Industrial Science and Technology,
Tsukuba, Ibaraki 305-8568, Japan}

\address{$^{4}$ JILA, University of Colorado and NIST, Boulder, CO 80309-0440,
USA}

\ead{nicholas.plumb@psi.ch, daniel.dessau@colorado.edu}
\begin{abstract}
Ultrahigh resolution angle-resolved photoemission spectroscopy (ARPES)
with low-energy photons is used to study the detailed momentum dependence
of the well-known nodal ``kink'' dispersion anomaly of Bi$_{2}$Sr$_{2}$CaCu$_{2}$O$_{8+\delta}$.
We find that the kink's location transitions smoothly from a maximum
binding energy of about 65 meV at the node of the $d$-wave superconducting
gap to 55 meV roughly one-third of the way to the antinode. Meanwhile,
the self-energy spectrum corresponding to the kink dramatically sharpens
and intensifies beyond a critical point in momentum space. We discuss
the possible bosonic spectrum in energy and momentum space that can
couple to the $k$-space dispersion of the electronic kinks. 
\end{abstract}


\maketitle

\section{Introduction}

One of the defining characteristics of the electronic structure of
the high-$T_{c}$ cuprates is the presence of an especially prominent
anomaly, or ``kink'', in the electronic dispersion, which corresponds
to a strong feature in the complex electronic self-energy spectrum
$\Sigma(\boldsymbol{k},\omega)=\Sigma'(\boldsymbol{k},\omega)+i\Sigma''(\boldsymbol{k},\omega)$.
The origin of the kink --- whether it is due to interactions of the
electrons with bosons (particularly phonons \cite{Devereaux2004,Ruiz2009}
or magnetic excitations \cite{Manske2001,Inosov2007,Dahm2009}) or
some other phenomenon \cite{Byczuk2007} --- is still heavily debated.
Likewise the kink's connection to superconductivity, and whether the
interactions it signifies may either form or break Cooper pairs, or
be altogether irrelevant, remains unknown.

At the nodes of the $d$-wave superconducting gap, this kink appears
at a binding energy of roughly 60--70 meV \cite{Bogdanov2000,Lanzara2001,Kaminski2001,Johnson2001}.
Meanwhile near the antinode, a seemingly stronger kink is located
at about 20--40 meV, depending on doping \cite{Campuzano1999,Gromko2003,Cuk2004}.
While a possible connection between the nodal and antinodal kinks
remains a mystery, the new data here fills in details of the evolving
physics between these points. Such information is crucial for obtaining
a complete understanding of the \foreignlanguage{british}{behaviour}
and origin of the kink and hence the electron-boson coupling in the
high-$T_{c}$ superconductors.

\section{Analysis and results}

\subsection{Experimental}

The data presented here were obtained from Bi$_{2}$Sr$_{2}$CaCu$_{2}$O$_{8+\delta}$
(Bi2212) near optimal doping with $T_{c}\approx89$ K. Rotational
alignment of the sample better than $1^{\circ}$ was performed by
Laue diffraction. The data were collected in the superconducting state
at 10 K using a photon energy of 7 eV. Compared to conventional photon
energies, the low photon energy greatly improves the photoelectron
escape depth, momentum resolution, and overall spectral sharpness
\cite{Koralek2006}. Total combined energy resolution of the light
source and \foreignlanguage{british}{analyser} was about 7 meV. ARPES
cuts were taken along the $(\pi,\pi)$ direction of the Fermi surface
(FS).

\subsection{Momentum-dependent self-energy}

In the present work, we are especially concerned with the self-energy
contribution due to electrons coupling to a collective mode over a
sharp energy range, and we wish to isolate this from other interactions
with smooth energy dependencies (e.g., electron-electron scattering
\cite{Ingle2005}). This is accomplished by assuming a smooth (in
this case linear) \textit{effective} bare band $\epsilon_{\text{eff}}(\boldsymbol{k})$
for each ARPES cut that connects points on the dispersion far from
the main kink. The real part of the effective self-energy is then
simply 
\begin{equation}
\Sigma'_{\text{eff}}(\omega)=\omega-v_{F}^{\text{eff}}[k_{m}(\omega)-k_{F}]\label{eq:eq1}
\end{equation}
where $k_{m}(\omega)$ is the measured dispersion, $v_{F}^{\text{eff}}$
is the slope of $\epsilon_{\text{eff}}(k)$, and $k_{F}$ is the Fermi
momentum. $\Sigma''_{\text{eff}}(\omega)$ is then the Kramers-Kronig
transformation of $\Sigma'_{\text{eff}}(\omega)$. Unlike $\Sigma'(\omega)$,
$\Sigma'_{\text{eff}}(\omega)$ is well-behaved at its endpoints (by
construction), and its Kramers-Kronig transformation is easily computed.
Our routine sets the in-gap points of $\Sigma'_{\text{eff}}(\omega)$
to zero and computes the transformation by Fast Fourier transform
assuming electron-hole symmetry. We have verified by simulations that
a possible violation of electron-hole symmetry \cite{Hashimoto2010}
should not significantly alter the findings here. We note that Eq.~\ref{eq:eq1}
is a conventional definition of $\Sigma'_{\text{eff}}$. Recently
it was shown that this definition undervalues the ``true'' bosonic
part of the self-energy by an overall scaling factor related to the
coupling strength of electron-electron interactions, $\lambda_{\text{el-el}}$
\cite{Iwasawa2013}. As this factor influences the magnitude of the
self-energy, not its distribution along $\omega$ or $\boldsymbol{k}$
, neglecting it will not affect the conclusions of the present study.
A systematic assessment of $\lambda_{\text{el-el}}$ in Bi2212, and
hence the correct scaling factor to be applied to $\Sigma'_{\text{eff}}$,
is currently underway.

\begin{figure}
\raggedleft\includegraphics[bb=0bp 0bp 389bp 520bp,width=0.85\textwidth]{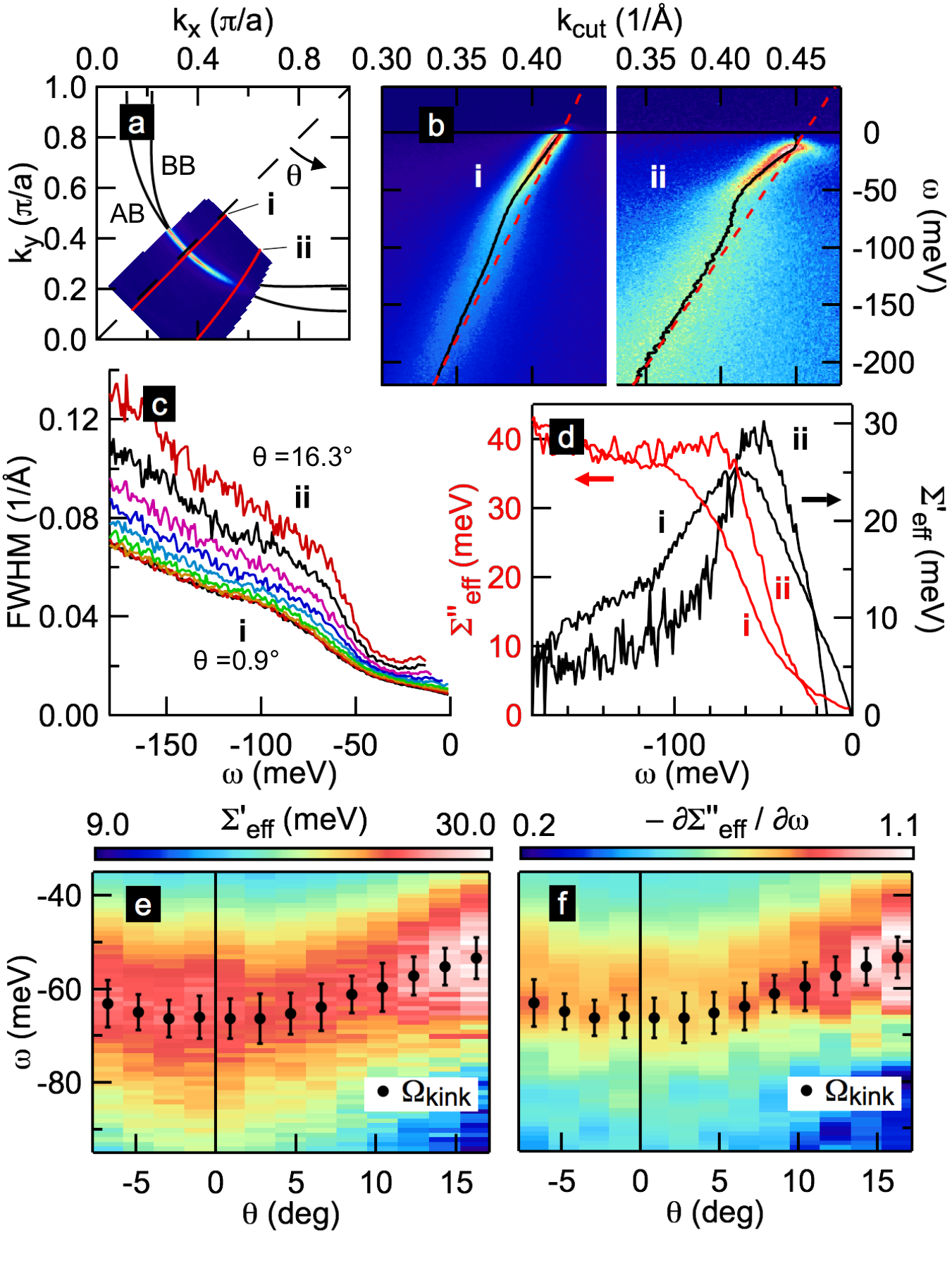}\caption{\label{fig:fig1} (a) First quadrant of the Fermi surface of Bi2212.
The \foreignlanguage{british}{colour} scale is the measured spectral
intensity 10 meV below $E_{F}$. Two representative cuts (\textbf{i}
and \textbf{ii}) are indicated by red curves. The black curves are
sketches of the antibonding (AB) and bonding band (BB) sheets. The
use of 7-eV photons isolates the antibonding band. (b) Raw ARPES data
from cuts \textbf{i} and \textbf{ii}. The solid black curves are the
peak positions of the fitted MDCs, while the dashed red lines are
the effective noninteracting bands for the dispersions (see text).
(c) MDC widths at cuts progressing away from the node. (d) Real (black)
and imaginary (red) components of the effective electronic self-energy
for cuts \textbf{i} and \textbf{ii}. (e) $\Sigma'_{\text{eff}}$ and
(f) $-\partial\Sigma''_{\text{eff}}/\partial\omega$ as a function
of $\theta$ and $\omega$. The black dots are the peak locations
of $\Sigma'_{\text{eff}}$ at each $\theta$, which we call $\Omega_{\text{kink}}(\theta)$.}
\end{figure}

Figure \ref{fig:fig1}(a) shows ARPES data collected along the FS
in the first quadrant of the Brillouin zone. The \foreignlanguage{british}{colour}
scale represents the measured intensity 10 meV below $E_{F}$. The
thick solid lines in Figure \ref{fig:fig1}(a) are sketches of the
antibonding (AB) and bonding band (BB) Fermi surfaces based on a tight-binding
model \cite{Markiewicz2004}. For 7-eV photons, only the AB is detected
\cite{Iwasawa2008}, which greatly simplifies the analysis. Two representative
raw data cuts, corresponding to Fermi surface angles $\theta=0.9^{\circ}$
and $\theta=16.3^{\circ}$, are indicated by the red curves labelled
\textbf{i} and \textbf{ii}, respectively. The spectra from these cuts
are shown in Figure \ref{fig:fig1}(b). The dispersions from momentum
distribution curve (MDC) fits are overlaid on the spectra (solid black
curves). The dashed red lines are assumed effective bare bands used
to calculate corresponding effective self-energy spectra $\Sigma_{\text{eff}}(\omega)$
at each $\theta$. These effective bare bands are determined by linear
fits from -230 meV to -200 meV that are constrained to pass through
the MDC peak location at $\omega=-\Delta(\theta)$ (i.e., $k_{F}$).

Figure \ref{fig:fig1}(c) shows the Lorentzian MDC widths for each
cut from \textbf{i} to \textbf{ii}, while Figure \ref{fig:fig1}(d)
depicts $\Sigma'_{\text{eff}}(\omega)$ (black, right axis) and $\Sigma''_{\text{eff}}(\omega)$
(red, left axis) for cuts \textbf{i} and \textbf{ii}. The full spectrum
of $\Sigma'_{\text{eff}}(\theta,\omega)$ is plotted as a \foreignlanguage{british}{colour}
scale in Figure \ref{fig:fig1}(e). We define the kink energy $\Omega_{\text{kink}}$
as the location of the peak in $\Sigma'_{\text{eff}}(\omega)$ at
each $\theta$. These values are determined by quadratic fits over
a range $\pm20$ meV about the maximum of each spectrum. The error
bars show the standard deviations ($\pm\sigma$) returned from the
fits. Figure \ref{fig:fig1}(f) depicts $-\partial\Sigma''_{\text{eff}}/\partial\omega$
as a function of $\omega$ and $\theta$. To reduce noise in the derivative,
some light smoothing was applied to the $\Sigma'_{\text{eff}}$ spectrum.
Together panels (e)-(f) highlight the evolution of the self-energy
over the nodal region, which exhibits both dispersive behaviour and
sharpening. The results are fully consistent with the behaviour of
the MDC widths in panel (c) and the electronic dispersion anomalies
in (b), providing an important verification of the self-consistency
of the data and analysis methods. It is worth noting that the quantity
$-\partial\Sigma''_{\text{eff}}/\partial\omega$ in panel (f) is somewhat
related to a useful parameter of strong coupling theory --- the \foreignlanguage{british}{Eliashberg}
boson coupling spectrum $\alpha^{2}F(\boldsymbol{k},\nu)$, where
$\nu$ is the energy axis for bosons. In an \textit{ungapped} system
at $T=0$, $\Sigma''(\boldsymbol{k},\omega)=\pi\int_{0}^{|\omega|}d\nu\alpha^{2}F(\boldsymbol{k},\nu)$
\cite{Mahan2000}, although the anisotropic gapping in cuprates can
significantly alter this relationship. Addressing this issue via suitable
``gap referencing'' is a key objective of the present work.

Two key points are evident from Figure \ref{fig:fig1}. First, $\Omega_{\text{kink}}$
evolves smoothly as a function of $\theta$ in the nodal region, shifting
toward $E_{F}$ by about 10 meV from $\theta=0$ to $\theta=15^{\circ}$.
Second, the nature of $\Sigma_{\text{eff}}$ appears to change abruptly
past a critical point in $\boldsymbol{k}$-space. While by eye the
kink perhaps becomes more dramatic going from node to antinode \cite{Sato2003},
this fact alone does not necessarily mean that the self-energy strengthens,
since $\Sigma$ is related to the bare band velocity, which decreases
away from the node. Indeed, the results in Figure \ref{fig:fig1}(c)-(f)
show that over much of the near-nodal region $\Sigma_{\text{eff}}(\omega)$
is relatively unchanged, despite the visual appearance that the kink
is ``getting stronger''. However, for $\theta\gtrsim10^{\circ}$
Figure \ref{fig:fig1}(e) shows a rapid increase in the peak of $\Sigma'_{\text{eff}}(\omega)$.
This corresponds with sharpening of the step in $\Sigma''_{\text{eff}}(\omega)$
seen in Figure \ref{fig:fig1}(f). The findings in Figure \ref{fig:fig1}
contrast with a previous study of overdoped Pb-Bi2212 where it was
argued that the scattering rate near $E_{F}$ is independent of $\boldsymbol{k}$
\cite{Bogdanov2002}. We also point out that these results contradict
recent claims that the energy $\Omega_{\text{kink}}$ is constant
near the node and then suddenly jumps at a ``crossover'' point on
the FS $\sim15^{\circ}$ away from the node \cite{Graf2008,Garcia2010},
though there still may be a crossover parameterised by, e.g., the
intensity and sharpness of the features in $\Sigma_{\text{eff}}(\omega,\theta)$,
as seen in Figure \ref{fig:fig1}(e)-(f).

\subsection{Scattering $\mathbf{q}$-space analysis of the kink momentum dependence}

The large nodal ARPES kink seen in cuprates is generally explained
as the result of the coupling of the electrons to a bosonic mode of
energy $\Omega_{\text{boson}}$. In particular, $\Omega_{\text{kink}}$
may be able to tell us which electrons interact with which bosons,
and in principle this can yield information about which (or even whether)
bosons act as the \textquotedblleft{}glue\textquotedblright{} responsible
for the formation of the Cooper pairs. The new finding of the large,
smooth dispersion of $\Omega_{\text{kink}}(\boldsymbol{k})$ is therefore
an important result that may connect directly to the coupling mechanism
of the electrons within a pair. Here we consider how to best connect
the $k$-dispersion of the kink to known data of the $q$-space dependence
of various bosonic modes. 

In the simplest picture, the kink energies $\Omega_{\text{kink}}$
will be exactly those of the coupling boson \cite{Lanzara2001}, though
this ignores the \textquotedblleft{}gap referencing\textquotedblright{}
which is simple for an $s$-wave superconductor ($|\Omega_{\text{kink}}|=\Omega_{\text{boson}}+\Delta$)
but more complicated for a $d$-wave superconductor in which $\Delta$
is strongly $k$-dependent. In the presence of an anisotropic gap
$\Delta(\boldsymbol{k})$, a bosonic mode with energy $\Omega_{\text{boson}}(\boldsymbol{q})$
scattering an electron purely from $\boldsymbol{k}$ to $\boldsymbol{k'}$
is expected to produce an ARPES dispersion kink below $E_{F}$ at
\cite{Plumb2010} 
\begin{equation}
|\Omega_{\text{kink}}(\boldsymbol{k})|=\Omega_{\text{boson}}(\boldsymbol{q})+\Delta(\boldsymbol{k'})\label{eq:eq2}
\end{equation}
which can be deduced by considering the set of photoholes at $\boldsymbol{k}$
that can be annihilated via electrons decaying from $\boldsymbol{k'}$
and emitting bosons $\Omega_{\text{boson}}(\boldsymbol{q})$. An argument
along these lines (but for an isotropic gap) is presented in section
7.3 of \cite{Mahan2000}. We will make use of this gap referencing
relationship throughout the present work in order to identify the
boson dispersions appropriate to particular scattering scenarios.
Additional corrections for relating $\Omega_{\text{kink}}$ to $\Omega_{\text{boson}}$
are believed to be too small to account for the dispersive behaviour
of $\Omega_{\text{kink}}(\boldsymbol{k})$ \cite{Schachinger2009a}
and therefore should not qualitatively alter the present work. 

\begin{figure}
\centering\includegraphics[height=0.6\textheight]{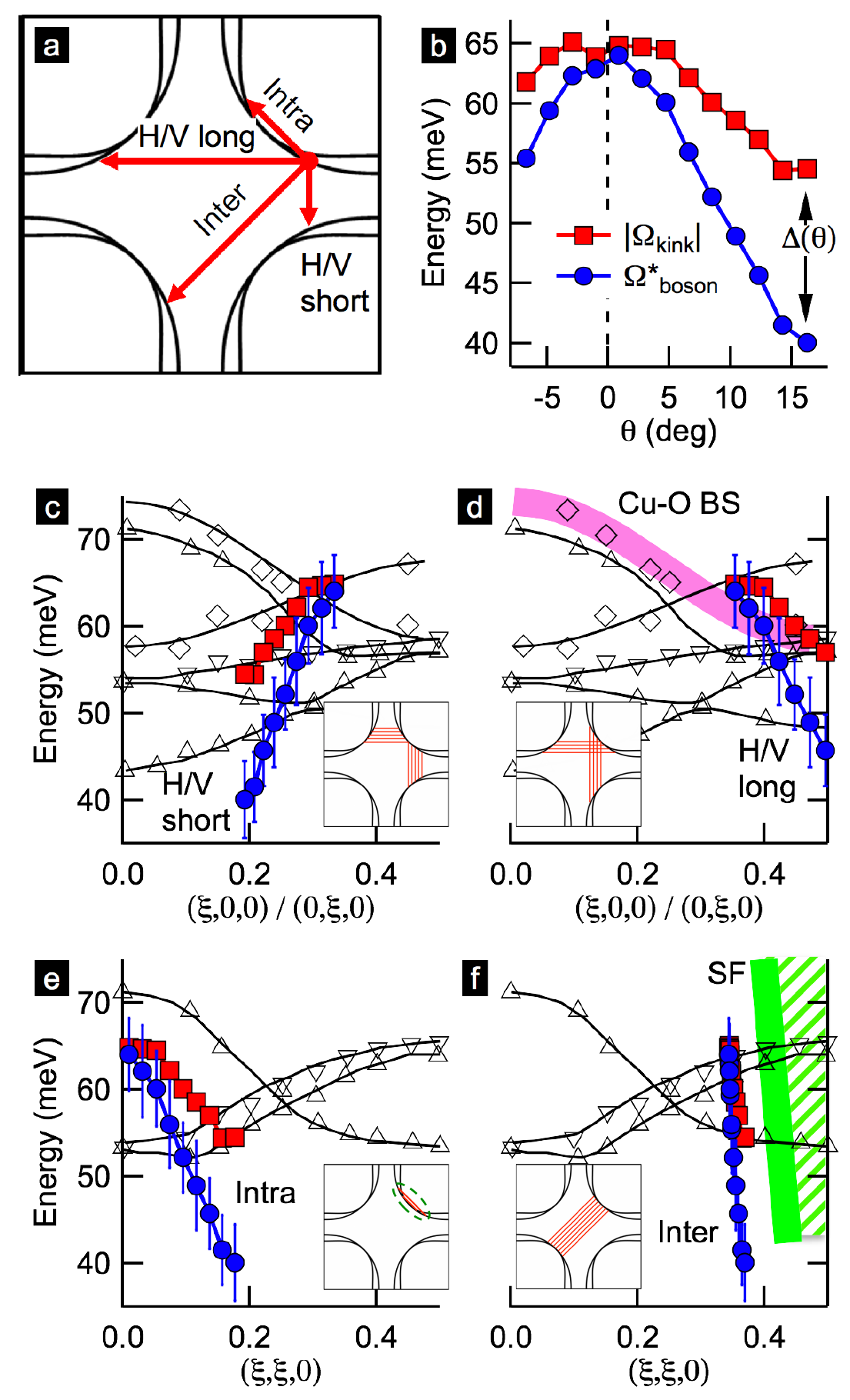}

\caption{\label{fig:fig2} Extracting the boson coupling mode dispersion assuming
the following scattering $\boldsymbol{q}$ directionalities in (a):
horizontal and vertical (H/V), intra-hole-pocket (Intra), and inter-hole-pocket
(Inter). (b) $|\Omega_{\text{kink}}|$ and $\Omega_{\text{boson}}^{*}$
as a function of FS angle $\theta$. $\Omega_{\text{boson}}^{*}$
{[}Equation (\ref{eq:eq3}){]} is the boson energy for the special
cases of scattering along symmetry directions such that $\Delta(\boldsymbol{k'})=\Delta(\boldsymbol{k})$,
as depicted in (a). (c), (d) Dispersions of $|\Omega_{\text{kink}}(\boldsymbol{q})|$
(red \fullsquare{}) and $\Omega_{\text{boson}}^{*}(\boldsymbol{q})$
(blue \fullcircle{}), assuming that scattering occurs horizontally/vertically
in the Brillouin zone. Curves for $|\Omega_{\text{kink}}(\boldsymbol{q})|$
and $\Omega_{\text{boson}}^{*}(\boldsymbol{q})$ have been extracted
considering the short (c) and long (d) H/V scattering channels separately.
The extracted dispersions are compared to those of Cu-O phonons measured
by INS in YBa\textsubscript{2}Cu\textsubscript{3}O\textsubscript{7}
(\opentriangle{}) and YBa\textsubscript{2}Cu\textsubscript{3}O\textsubscript{6}
(\opentriangledown{}) \cite{Reichardt1996}, as well as phonons observed
by IXS in Bi2201 (\opendiamond{}) \cite{Graf2008}. The pink highlighted
Cu-O bond-stretching (Cu-O BS) branch, in particular, has been identified
by some previous experiments as possibly relevant to the nodal kink.
(e), (f) Analogous plots assuming diagonal intra- and inter-pocket
scattering. The green shaded line in (f) is the approximate dispersion
of the high-energy branch of spin fluctuations (SF) observed in optimally-doped
Bi2212 \cite{Xu2009}. The hatched area indicates that this region
is observed in the neutron data to be somewhat filled in by the width
of the dispersion peaks. For simplicity, error bars from Figures \ref{fig:fig1}(e)
and (f) are shown only for the $\Omega_{\text{boson}}^{*}(\boldsymbol{q})$
curve in each panel.}
\end{figure}

There is reason to believe that the predominant electron-boson scattering
relevant to the nodal kink falls along some symmetry direction, thus
simplifying the connection between $k$- and $q$-space. For instance,
spin fluctuations observed by inelastic neutron scattering (INS) are
peaked at points on or near the $(\xi,\xi,0)$ line \cite{Pailhes2004,Hayden2004,Tranquada2004}.
Likewise, both experiment \cite{Reznik2010} and theory \cite{Devereaux2004,Giustino2008}
suggest the Cu-O ``half-breathing'' phonon mode scatters electrons
primarily along $(\xi,0,0)$ {[}equivalently $(0,\xi,0)${]}, and
there is evidence that this mode couples strongly to electrons \cite{Pintschovius2005}
and, in particular, may contribute to the nodal kink \cite{Iwasawa2008}.

In the case of phonons, numerical calculations find that, on the whole,
the scattering matrix elements are fairly complicated \cite{Devereaux2004,Giustino2008}.
Nevertheless, they are expected to evolve smoothly over the FS and
exhibit some preference for particular directionalities. Thus, despite
the complexity of the full scattering problem, one can reasonably
expect to find \textit{qualitative} agreement between the actual phonon
dispersion and the inference from ARPES --- at least over a limited
portion of the FS.

To proceed with our analysis, assumed scattering $\boldsymbol{q}$'s
along symmetry directions are illustrated in Figure \ref{fig:fig2}(a).
We consider cases where electrons may scatter horizontally, vertically,
or diagonally via inter- or intra-hole-pocket vectors. The kink energies
obtained in Figure \ref{fig:fig1}(e) are plotted in Figure \ref{fig:fig2}(b)
as a function of FS angle $\theta$ (red \fullsquare{}). The blue
circles (\fullcircle{}) are the corresponding gap-referenced boson
energies $\Omega_{\text{boson}}^{*}$. The asterisk ({*}) denotes
that only the special cases of scattering vectors shown in that panel
apply. Under these circumstances, $\Delta(\boldsymbol{k})=\Delta(\boldsymbol{k'})$,
leading to 
\begin{equation}
\Omega_{\text{boson}}^{*}(\theta)=|\Omega_{\text{kink}}(\theta)|+\Delta(\theta)\text{.}\label{eq:eq3}
\end{equation}
In calculating $\Omega_{\text{boson}}^{*}(\theta)$, we used $\Delta(\theta)$
based on our ARPES-measured values, which were found to have excellent
agreement with the expected $d$-wave form, with maximum (antinodal)
magnitude $\Delta_{0}=30$ meV. The gap measurements shown here were
performed using the newly-developed tomographic density of states
(TDoS) technique \cite{Reber2012,Reber2013}, and we have checked
that the analysis and results that follow are essentially unchanged
if the symmetrised energy distribution curve (EDC) method is employed
\cite{Norman1998}.

The extracted $\boldsymbol{q}$-space dispersions of $\Omega_{\text{kink}}$
and $\Omega_{\text{boson}}^{*}$ under these various scattering scenarios
are shown in Figure \ref{fig:fig2}(c)--(f). The insets in each of
these panels illustrate how the $q$ values on each horizontal axis
were determined. For simplicity and generalizability of the analysis,
we consider each scattering channel independently. For an assumed
bosonic mode that would scatter electrons in the $(\xi,0,0)/(0,\xi,0)$
directions, a given $\boldsymbol{k}$ point on the Fermi surface could
couple via two orthogonal vectors --- one shorter than the node-node
distance in $\boldsymbol{q}$-space ($\xi\sim0.35$) and the other
longer {[}labelled ``H/V short'' and ``H/V long'' in Figure \ref{fig:fig2}(c)
and (d), respectively{]}. However, in general these short and long
$\boldsymbol{q}$ channels would not be expected to contribute equally
to the appearance of the kink, but rather their relative scattering
intensities would evolve around the Fermi surface (only matching at
the node-node distance, where they have the same length). To disentangle
these paired interactions, the $\Omega_{\text{kink}}(\boldsymbol{q})$
and $\Omega_{\text{boson}}^{*}(\boldsymbol{q})$ curves were extracted
by treating the short and long scattering vectors separately, as plotted
in Figure \ref{fig:fig2}(c), (d). This is a logical choice, since
one or the other (either the short or long vector) would probably
be more influential at any given point on the Fermi surface. Meanwhile,
the diagonal intra- and inter-hole-pocket vectors {[}``Intra'' and
``Inter'' in Figure \ref{fig:fig2}(e) and (f) respectively{]} are
also treated independently, since they are distinct in how they couple
the topology of the Fermi surface.

In Figure \ref{fig:fig2}(c)--(f), the extracted $\Omega_{\text{boson}}^{*}(\boldsymbol{q})$
curves (blue \fullcircle{}) are compared to various Cu-O phonon dispersions
in YBa$_{2}$Cu$_{3}$O$_{6+x}$ (YBCO) with $x=0$ (\opentriangledown{})
and $x=1$ (\opentriangle{}) \cite{Reichardt1996}, as well as two
phonon branches observed in Bi$_{2}$Sr$_{1.6}$La$_{0.4}$Cu$_{2}$O$_{6+\delta}$
(Bi2201, \opendiamond{}) \cite{Graf2008}. Additionally, a sketch
of the dispersion of a high-energy branch of incommensurate spin fluctuations
is shown in Figure \ref{fig:fig2}(f) (green shaded line).

Overall, the extracted $\Omega_{\text{boson}}^{*}(\boldsymbol{q})$
curves do not provide clear support that the kink primarily originates
from electron-phonon interactions, although the data may not be wholly
inconsistent with this possibility. For instance, in Figure 2(d),
there is a limited $\boldsymbol{q}$-space region {[}$\boldsymbol{q}\approx(0.3\text{--}0.4,0,0)${]}
where the extracted boson dispersion roughly overlaps with the Cu-O
bond stretching phonon branch, as pointed out in \cite{Graf2008}.
However, for larger values of $\boldsymbol{q}$ extending toward $(0.5,0,0)$,
$\Omega_{\text{boson}}^{*}(\boldsymbol{q})$ diverges from the Cu-O
bond stretching phonon branch with a different slope. In this regard,
our data show greater overall similarity between $\Omega_{\text{boson}}^{*}(\boldsymbol{q})$
and the SF dispersion in Figure \ref{fig:fig2}(f), which differ by
merely a simple offset in $\omega$ and/or $\boldsymbol{q}$, perhaps
reflecting systematic differences between the techniques and/or samples
used in the studies. The rough correspondence between the kink and
the SF dispersion compares favorably with spectral analysis of the
spin response function extracted from ARPES data by Chatterjee \etal~\cite{Chatterjee2007},
though their formalism only considered spin fluctuations and did not
provide a comparison to phonon dispersions. Moreover, that approach
treated the spectral function holistically, rather than isolating
the kink feature and considering its explicit connection to the bosonic
spectral function.

\begin{figure}
\raggedleft\includegraphics[bb=0bp 5bp 334bp 177bp,width=0.85\textwidth]{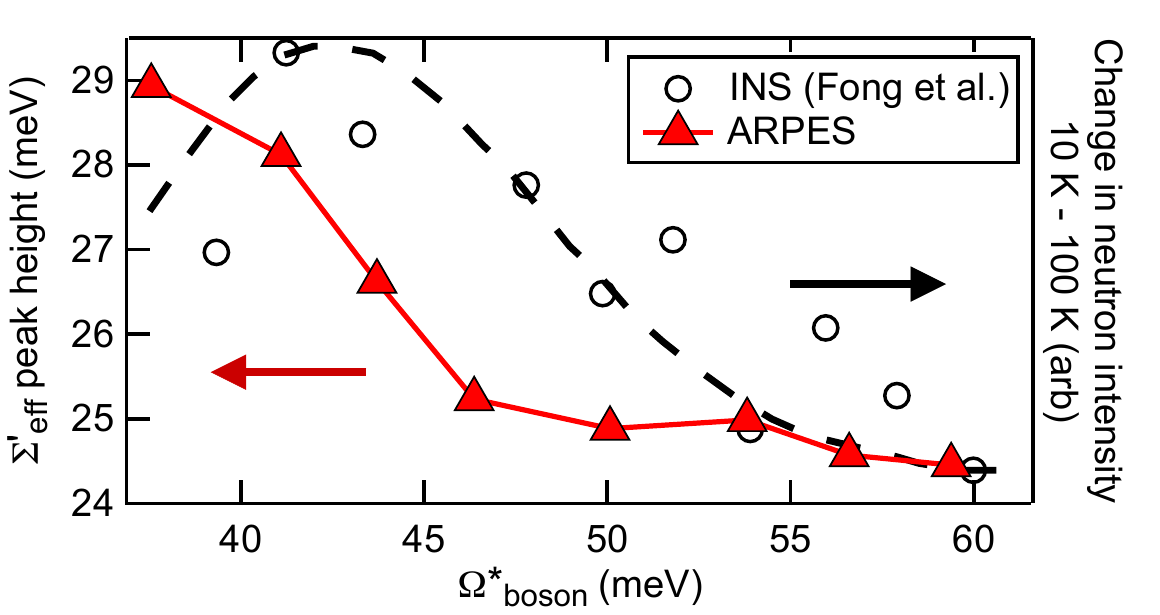}

\caption{\label{fig:fig3} Peak height of $\Sigma'_{\text{eff}}$ as a function
of $\Omega_{\text{boson}}^{*}$. The results are compared to INS data
\cite{Fong1999} highlighting the magnetic resonance at $\sim40$
meV. The INS data points are the neutron scattering intensity at 100
K subtracted from the signal at 10 K.}
\end{figure}

The SF dispersion in Figure \ref{fig:fig2}(f) is the high-energy
branch of incommensurate spin excitations so far observed in many
cuprates \cite{Pailhes2004,Hayden2004,Reznik2004,Tranquada2004,Xu2009}.
It converges with a low-energy branch near $\sim40$ meV, where there
is a well-known $\boldsymbol{q}=(0.5,0.5)$-centred ``resonance''
peak in the spin susceptibility at low $T$ \cite{Rossat-Mignod1991,Mook1993,Dai1998,Fong1999}.
The effective self-energy obtained from our analysis intensifies significantly
at $\Omega_{\text{boson}}^{*}$ somewhat near the resonance energy.
This is depicted in Figure \ref{fig:fig3}, where the peak height
of $\Sigma'_{\text{eff}}$ (red triangles) is plotted versus $\Omega_{\text{boson}}^{*}$.
INS data from optimally-doped Bi2212 (open circles) show the difference
in scattered neutron intensity from 100 K to 10 K, illustrating the
location of the resonance \cite{Fong1999}. Notably, within the context
of an orbital-overlap model, coupling to the Cu-O bond-stretching
phonon suggested by Figure \ref{fig:fig2}(d) is not expected to intensify
in this manner at $\theta$ corresponding to $\Omega_{\text{boson}}^{*}$
close to the resonance \cite{Johnston2010a}. With that said, the
data is again only in rough agreement with the SF picture, and other
studies imply that the strength of $\Sigma_{\text{eff}}^{'}$ is monotonic
around the Fermi surface \cite{Terashima2006}, meaning that it would
not obey the peak-shaped trend of the INS data reproduced in Figure
\ref{fig:fig3}. However, this does not fully undermine the possibility
that the kink is SF-related, as it could instead signal a contribution
to $\Sigma_{\text{eff}}^{'}$ at low energies (i.e., near the antinode)
from additional types of electron-boson interactions, as we will discuss. 

\section{Discussion}

The above analysis has relied on the key assumption of a dominant
scattering mode and directionality, which, while consistent with interpretations
of some INS and inelastic x-ray scattering (IXS) data and calculations,
is not exhaustive. A natural potential counter example is the case
where all points on the FS couple primarily to the van Hove singularities
at the antinodes --- a situation in which the results would not be
directly comparable with conventional INS/IXS data, since the $\boldsymbol{q}$'s
would no longer fall along a straight line. Perhaps the best that
can be said is that coupling strictly to the antinodes would shift
the $\Omega_{\text{kink}}(\boldsymbol{q})$ dispersion down universally
by $\Delta_{0}\approx30$ meV such that $\Omega_{\text{boson}}(\boldsymbol{q})$
would span an energy range of roughly 25--35 meV. These energies are
home to many phonons in the cuprates \cite{Falter2005}, which in
principle could combine their effects in some intricate way to produce
the observed kink behaviour. A strong antinodal coupling appears unlikely,
however, because this scenario would imply a huge shift of the nodal
kink energy between the normal and superconducting states and/or as
a function of doping. Multiple ARPES studies find no evidence of such
a shift \cite{Johnson2001,Gromko2003,Sato2003,Zhang2008,Plumb2010},
although one case where the kink was interpreted to be composed of
multiple phonon mode couplings arguably shows evidence of node-antinode
scattering \cite{Lee2008}.

Reviewing the results, the analysis of $\Omega_{\text{boson}}^{*}(\boldsymbol{q})$
for ``long'' $(\xi,0,0)$/$(0,\xi,0)$ scattering vectors found
some region of agreement with the dispersion of a Cu-O bond stretching
phonon {[}Figure \ref{fig:fig2}(d){]}, although the trends diverge
as $\xi$ extends out to 0.5. On the other hand, $\Omega_{\text{boson}}^{*}(\boldsymbol{q})$
extracted for diagonal inter-hole-pocket scattering is similar to
the dispersion of spin fluctuations {[}Figure \ref{fig:fig2}(f){]},
merely differing by a simple offset in energy and/or $\boldsymbol{q}$.
Additionally, Figure \ref{fig:fig3} illustrates that the strength
of the self-energy associated with the kink, plotted with respect
to $\Omega_{\text{boson}}^{*}$, has a qualitative resemblance to
spin fluctuations, and this is contrary to the behaviour predicted
for electron-phonon coupling \cite{Johnston2010a}. However, to the
extent the data might be viewed as favoring the spin fluctuation picture,
it poses an intriguing apparent paradox; A very detailed low-$h\nu$
ARPES study found an isotope shift in the energy of the nodal kink
\cite{Iwasawa2008}, giving strong merit to the phonon scenario. One
possible explanation for this conflict is that electron-phonon interactions
might constitute a finite but relatively small contribution to the
total self-energy \cite{Schachinger2009}. Alternatively, the results
may signal a role for coupling between spin and lattice degrees of
freedom \cite{Nazarenko1996,Normand1996,Nunner1999}. 

The discovery of the large momentum dependence of the main nodal kink
adds to the richness of strong electron-boson coupling phenomena in
cuprates. It was recently shown that a newly-discovered ultra-low-energy
kink $\sim10$ meV below $E_{F}$ \cite{Rameau2009,Plumb2010,Vishik2010,Anzai2010}
has its own distinct momentum dependence that runs counter to the
behaviour of the deeper-energy kink studied here \cite{Johnston2012}.
Specifically, unlike the larger main kink 65--55 meV below $E_{F}$,
which evolves toward lower binding energy while moving from node to
antinode, the ultra-low-energy kink closely follows the contour of
the superconducting gap in the nodal region, moving to higher binding
energy approaching the antinode. A natural, but spectroscopically
demanding, next course of study will be to investigate the possible
convergence of these two energy scales near the antinodal point and
to see whether either or both this these connect with the antinodal
feature observed near 20--40 meV, which historically has been regarded
as a separate kink. Hence the data here, in concert with \cite{Johnston2012},
open the possibility that interactions with distinct physical origins
could combine within a narrow energy range in the antinodal region
--- perhaps with major implications for high-$T_{c}$ superconductivity.

In conclusion, using low photon energy ARPES, we have mapped the detailed
momentum dependence of the primary kink in the nodal $k$-space region
of near-optimal Bi2212. From a simplifying treatment of the data that
takes into account effects of the $d$-wave superconducting gap, the
kink's dispersion seems inconsistent with most phonons, though over
a limited range of momentum transfer {[}$\boldsymbol{q}\approx(0.3\text{--}0.4,0,0)${]}
it bears some semblance to scattering due to a Cu-O bond-stretching
mode. However, in terms of the momentum dependence of the location
and sharpness/intensity of the self-energy feature, the greatest similarity
is found with the dispersion of the upper branch of incommensurate
spin fluctuations.

\emph{Note added} --- During review of this manuscript, a related
article was published \cite{He2013a}.

\ack{}{Funding was provided by the DOE under project number DE-FG02-03ER46066.
Experiments were conducted at BL-9A of the Hiroshima Synchrotron Radiation
Center and BL5-4 of the Stanford Synchrotron Radiation \foreignlanguage{british}{Lightsource}
(SSRL). SSRL is operated by the DOE, Office of Basic Energy Sciences.
We thank D. Reznik, T. P. Devereaux, and S. Johnston for valuable
conversations.}

\bibliographystyle{unsrt}
\bibliography{citations}

\end{document}